\documentclass[preprint]{ptephy_v1}

\preprintnumber{KOBE-TH-19-01, KYUSHU-HET-193} 

\usepackage{amsmath} 

\newcommand{\vev}[1]{\left\langle #1 \right\rangle}
\newcommand{\vvev}[1]{\left\langle\kern-0.3em\left\langle #1
      \right\rangle\kern-0.3em\right\rangle} 
\newcommand{\lb}{\left\lbrace}
\newcommand{\rb}{\right\rbrace}
\newcommand{\SL}{S_\Lambda}
\newcommand{\Op}{\mathcal{O}}
\newcommand{\Sb}{S_{\mathrm{bare}}}

\begin{document}

\title{Derivation of a Gradient Flow from ERG}


\author[1]{Hidenori~Sonoda}
\affil{Physics Department, Kobe University, Kobe 657-8501, Japan\email{hsonoda@kobe-u.ac.jp}}

\author[2]{Hiroshi~Suzuki}
\affil{Department of Physics, Kyushu University, 744 Motooka,
  Nishi-Ku, Fukuoka 819-0395, Japan\email{hsuzuki@phys.kyushu-u.ac.jp}}

\begin{abstract}%
  We establish a concrete correspondence between a gradient flow and
  the renormalization group flow for a generic scalar field theory.
  We use the exact renormalization group formalism with a particular
  choice of the cutoff function.
\end{abstract}

\subjectindex{}

\maketitle

\section{Introduction}

The gradient flow, introduced in~\cite{Narayanan:2006rf,Luscher:2010iy}, has
been attracting much attention lately.  It is a continuous diffusion of local
fields, well defined not only in continuum space but also on discrete lattices.
The flow is much reminiscent of the renormalization group
transformation~\cite{Wilson:1973jj}, and it is especially important for lattice
theories which had only discrete renormalization group transformations
available.

The gradient flow has been used for the scale setting and for the definition of
the topological charge~\cite{Luscher:2010iy} (see \cite{Ramos:2015dla} for a
review). It also has been used to compute the expectation values of physical
quantities such as the energy-momentum tensor via the small diffusion time
expansions~\cite{Luscher:2011bx} of local products of
fields~\cite{Suzuki:2013gza,Makino:2014taa}. It has also been used to find
non-trivial infrared behaviors of theories such as scalar theories with O($N$)
invariance and QCD with massless quarks; see, for
instance~\cite{Carosso:2018bmz,Carosso:2018rep}.

The similarity between the gradient flow and the renormalization group
flow was pointed out already at the beginning~\cite{Luscher:2010iy}
and has been pursued
further~\cite{Luscher:2013vga,Kagimura:2015via,Yamamura:2015kva,%
  Aoki:2016ohw,Makino:2018rys,Abe:2018zdc,Carosso:2018bmz,Carosso:2018rep}. The
purpose of this paper is to establish a concrete correspondence
between the two flows for a generic real scalar field theory in
$D$-dimensional Euclidean space.  We introduce Wilson actions with a
finite momentum cutoff using the formalism of the exact
renormalization group (ERG)~\cite{Wilson:1973jj}. Many readers may not
have sufficient familiarity with the formalism, and we have chosen to
give ample background material with the expense of brevity of the
paper.

In the gradient flow we introduce a diffusion time $t > 0$, and extend
the local field $\phi (x)$ in $D$-dimensional space by the solution of
the diffusion equation (with no non-linear terms; see \cite{Capponi:2015ucc}
for the motivation for this simple choice):
\begin{equation}
\partial_t \varphi (t,x) = \partial^2 \varphi (t,x)\,,
\end{equation}
where $\varphi (0,x) = \phi (x)$.  We will show that the correlation
functions of the diffused field $\varphi (t,x)$ evaluated in the bare
action $\Sb$ of the theory are given by the correlation functions of
the elementary field $\phi (x)$ evaluated in the Wilson action $\SL$
with a finite momentum cutoff $\Lambda$:
\begin{equation}
  (\varphi (t,x), \Sb) \Longleftrightarrow (\phi (x),
  S_\Lambda)\,,\label{correspondence} 
\end{equation}
where $t$ and $\Lambda$ are related by
\begin{equation}
t = \frac{1}{\Lambda^2}\,.
\end{equation}
Based on this correspondence (\ref{correspondence}), the renormalized
nature of the diffused field results from the finite momentum cutoff
of the Wilson action.

We organize the paper as follows.  In Sec. 2 we briefly overview the
ERG formalism.  We provide more as we proceed.  In Sec. 3 we derive a
gradient flow from ERG.  We consider a generic scalar theory, not
necessarily renormalizable, and consider the behavior of the gradient
flow at large diffusion times.  In Sec. 4 we extend the gradient flow
to renormalizable theories.  We follow Sec. 12 of~\cite{Wilson:1973jj}
to renormalize a theory non-perturbatively.  This is to prepare for
the discussion of the gradient flow at small diffusion times in
Sec. 5, where we derive the small time expansions of local products of
the diffused field.  In Sec. 6 we conclude the paper.

We use the following shorthand notation for the momentum integrals:
\begin{equation}
\int_p \equiv \int \frac{d^D p}{(2 \pi)^D},\quad
\delta (p) \equiv (2 \pi)^D \delta^{(D)} (p)\,.
\end{equation}

\section{Overview of the ERG formalism}

We give a brief overview of the exact renormalization group.  There
are many reviews available on the subject (\cite{Igarashi:2009tj} and
references therein); we follow the convention of~\cite{Sonoda:2015bla}
in the following.

Let $\SL [\phi]$ be a Wilson action of a real scalar field with a
momentum cutoff $\Lambda$.  The cutoff dependence of the Wilson action
is determined so that the physics contents remain unchanged.  We use
the convention that the Boltzmann factor of functional integration is
$e^{\SL [\phi]}$ rather than the more common $e^{- \SL [\phi]}$.  The
$\Lambda$ dependence is given by the ERG differential equation in
momentum space:
\begin{align}
  - \Lambda \frac{\partial}{\partial \Lambda} e^{\SL [\phi]} &= \int_p
  \left[ \left( \frac{\Delta (p/\Lambda)}{K(p/\Lambda)} -
      \frac{\eta}{2}
    \right) \phi (p) \frac{\delta}{\delta \phi (p)}\right.\notag\\
  &\quad\left. + \frac{\Delta (p/\Lambda) - \eta K(p/\Lambda) \left( 1
        - K(p/\Lambda) \right)}{p^2} \frac{1}{2}
    \frac{\delta^2}{\delta \phi (p) \delta \phi (-p)} \right]\, e^{\SL
    [\phi]}\,,\label{ERGdiffeq}
\end{align}
where 
$K(p/\Lambda)$ is a positive cutoff function that decreases
rapidly as $p \to \infty$, and
\begin{equation}
\Delta (p/\Lambda) \equiv \Lambda \frac{\partial}{\partial \Lambda}
K(p/\Lambda)\,.
\end{equation}
We have introduced a constant anomalous dimension $\eta > 0$.  Any
cutoff function will do as long as
\begin{equation}
K(0) = 1,\quad \lim_{p \to \infty} K(p/\Lambda) = 0\,.
\end{equation}
In this paper we choose
\begin{equation}
K(p/\Lambda) = e^{- p^2/\Lambda^2}\label{gaussian}
\end{equation}
so that the inverse squared cutoff $1/\Lambda^2$ can play the role of
a diffusion time $t$ for the gradient flow.

Given a bare action $\Sb [\phi]$ at $\Lambda_0$, we can solve
(\ref{ERGdiffeq}) for $\Lambda < \Lambda_0$ by an integral
formula~\cite{Sonoda:2015bla}
\begin{align}
  &e^{\SL [\phi]} = \int [d\phi'] \exp \left[ \Sb [\phi'] -
    \frac{1}{2} \int_p
    \frac{p^2}{\left(\frac{\Lambda}{\mu}\right)^\eta
      \frac{1-K(p/\Lambda)}{K(p/\Lambda)} -
      \left(\frac{\Lambda_0}{\mu}\right)^\eta \frac{1-
        K(p/\Lambda_0)}{K(p/\Lambda_0)}}\right.\notag\\
  &\left. \,\times \left(
      \left(\frac{\Lambda_0}{\mu}\right)^{\frac{\eta}{2}} \frac{\phi'
        (p)}{K(p/\Lambda_0)} -
      \left(\frac{\Lambda}{\mu}\right)^{\frac{\eta}{2}} \frac{\phi
        (p)}{K(p/\Lambda)} \right) \left(
      \left(\frac{\Lambda_0}{\mu}\right)^{\frac{\eta}{2}} \frac{\phi'
        (-p)}{K(p/\Lambda_0)} -
      \left(\frac{\Lambda}{\mu}\right)^{\frac{\eta}{2}} \frac{\phi
        (-p)}{K(p/\Lambda)} \right) \right]\,.\label{integral}
\end{align}
The dependence on the reference momentum $\mu$ is only apparent.  In
the next section we use this formula to relate the correlation
functions of $S_\Lambda$ to the bare correlation functions.

As it is, the ERG differential equation (\ref{ERGdiffeq}) has no fixed
point.  To obtain an ERG differential equation with a fixed point, we
must measure dimensionful quantities in units of appropriate powers of
the cutoff $\Lambda$.  We introduce a dimensionless field with a
dimensionless momentum by
\begin{align}
  \bar{p} &\equiv \frac{p}{\Lambda}\,,\\
  \bar{\phi} (\bar{p}) &\equiv \Lambda^{\frac{D+2}{2}} \phi (\bar{p}
  \Lambda)\,.
\end{align}
Defining $\tau \equiv \ln \frac{\mu}{\Lambda}$, we can rewrite
(\ref{ERGdiffeq}) for
\begin{equation}
\bar{S}_\tau [\bar{\phi}] \equiv \SL [\phi]
\end{equation}
as follows:
\begin{align}
  \partial_\tau e^{\bar{S}_\tau [\bar{\phi}]} &= \int_{\bar{p}} \left[
    \left( \frac{\Delta (\bar{p})}{K(\bar{p})} + \frac{D+2}{2} -
      \frac{\eta}{2} + \bar{p} \cdot \partial_{\bar{p}} \right)
    \bar{\phi} (\bar{p}) \frac{\delta}{\delta
      \bar{\phi} (\bar{p})} \right.\notag\\
  &\left.\quad + \frac{\Delta (\bar{p}) - \eta K(\bar{p}) \left( 1 -
        K(\bar{p})\right)}{\bar{p}^2} \frac{1}{2}
    \frac{\delta^2}{\delta \bar{\phi} (\bar{p}) \delta \bar{\phi} (-
      \bar{p})} \right] \, e^{\bar{S}_\tau
    [\bar{\phi}]}\,.\label{ERGdiffeq-Sbar}
\end{align}
With an appropriate choice of $\eta$, this can have a non-trivial
fixed point action $\bar{S}^*$ for which the right-hand side above
vanishes.

\section{Derivation of a gradient flow}

To derive a gradient flow for the scalar field, we need to rewrite
(\ref{integral}) for $\bar{S}_\tau [\bar{\phi}]$.  The calculation is
straightforward, and we just write down the result:
\begin{align}
  e^{\bar{S}_\tau [\bar{\phi}]} &= \int [d\phi'] \exp \left[ \Sb
    [\phi'] - \frac{1}{2} \int_{\bar{p}} \frac{\bar{p}^2}{\frac{1 -
        K(\bar{p})}{K(\bar{p})} -
      \left(\frac{\Lambda_0}{\Lambda}\right)^\eta \frac{1-K\left(
          \bar{p} \frac{\Lambda}{\Lambda_0}\right)}{K\left(\bar{p}
          \frac{\Lambda}{\Lambda_0}\right)}}
  \right.\notag\\
  &\left.\times \left(
      \left(\frac{\Lambda_0}{\Lambda}\right)^{\frac{\eta}{2}}
      \frac{\Lambda^{\frac{D+2}{2}} \phi' (\bar{p}
        \Lambda)}{K\left(\bar{p} \frac{\Lambda}{\Lambda_0}\right)} -
      \frac{\bar{\phi} (\bar{p})}{K(\bar{p})}\right) \left(
      \left(\frac{\Lambda_0}{\Lambda}\right)^{\frac{\eta}{2}}
      \frac{\Lambda^{\frac{D+2}{2}} \phi' (- \bar{p}
        \Lambda)}{K\left(\bar{p} \frac{\Lambda}{\Lambda_0}\right)} -
      \frac{\bar{\phi} (-\bar{p})}{K(\bar{p})}\right) \right]\,.
\label{Sbar-Sbare}
\end{align}
We introduce the generating functionals for $\bar{S}_\tau$ and $\Sb$:
\begin{align}
  \bar{Z}_\tau [\bar{J}] &\equiv \int [d\bar{\phi}] \exp \left[
    \bar{S}_\tau [\bar{\phi}] + \int_{\bar{p}} \bar{J} (-\bar{p})
    \bar{\phi} (\bar{p}) \right]\,,\label{Zbar-def}\\
  Z_0 [J'] &\equiv \int [d \phi'] \exp \left[ \Sb [\phi'] + \int_p J'
    (-p) \phi' (p) \right]\,.\label{Zbare-def}
\end{align}
By substituting (\ref{Sbar-Sbare}) into (\ref{Zbar-def}), and integrating
over $\bar{\phi}$ first, we obtain~\cite{Igarashi:2009tj}
\begin{align}
  \bar{Z}_\tau [\bar{J}] &= \int [d\phi'] \exp \left[ \Sb [\phi'] +
    \int_{\bar{p}} \bar{J} (- \bar{p}) K(\bar{p})
    \left(\frac{\Lambda_0}{\Lambda}\right)^{\frac{\eta}{2}}
    \frac{\Lambda^{\frac{D+2}{2}} \phi' (\bar{p} \Lambda)}{K
      \left(\bar{p}
        \frac{\Lambda}{\Lambda_0}\right)}\right.\notag\\
  &\left. \quad + \frac{1}{2} \int_{\bar{p}} \bar{J} (\bar{p}) \bar{J}
    (- \bar{p}) \frac{K(\bar{p})^2}{\bar{p}^2} \lb
    \frac{1-K(\bar{p})}{K(\bar{p})} -
    \left(\frac{\Lambda_0}{\Lambda}\right)^\eta \frac{1 - K
      \left(\bar{p} \frac{\Lambda}{\Lambda_0}\right)}{K \left(\bar{p}
        \frac{\Lambda}{\Lambda_0}\right)} \rb \right]\notag\\
  &= Z_0 \left[ J' \right] \exp \left[ \frac{1}{2} \int_{\bar{p}}
    \bar{J} (\bar{p}) \bar{J} (- \bar{p})
    \frac{K(\bar{p})^2}{\bar{p}^2} \lb \frac{1-K(\bar{p})}{K(\bar{p})}
    - \left(\frac{\Lambda_0}{\Lambda}\right)^\eta \frac{1 - K
      \left(\bar{p} \frac{\Lambda}{\Lambda_0}\right)}{K \left(\bar{p}
        \frac{\Lambda}{\Lambda_0}\right)} \rb
  \right]\,,\label{Zbar-Zbare}
\end{align}
where $J'$ is given by
\begin{equation}
  J' (\bar{p} \Lambda) \equiv \bar{J} (\bar{p}) \frac{K(\bar{p})}{K
    \left( \bar{p} \frac{\Lambda}{\Lambda_0}\right)}
  \left(\frac{\Lambda_0}{\Lambda}\right)^{\frac{\eta}{2}} \Lambda^{-
    \frac{D-2}{2}} \,. 
\end{equation} 
The extra quadratic term in $\bar{J}$'s only affect the two-point function.

The result (\ref{Zbar-Zbare}) implies that the two-point function of
$\bar{S}_\tau$ differs from that of $\Sb$ by normalization and a shift,
both momentum dependent:
\begin{align}
  \vev{\bar{\phi} (\bar{p}) \bar{\phi} (\bar{q})}_{\bar{S}_\tau} &=
  \left(\frac{\Lambda_0}{\Lambda}\right)^\eta \Lambda^{D+2}
  \left(\frac{K(\bar{p})}{K\left(\bar{p}
        \frac{\Lambda}{\Lambda_0}\right)}\right)^2
  \vev{\phi (\bar{p} \Lambda) \phi (\bar{q} \Lambda)}_{\Sb}\notag\\
  &\quad + \delta (\bar{p} + \bar{q}) \frac{K(\bar{p})^2}{\bar{p}^2}
  \lb \frac{1 - K(\bar{p})}{K(\bar{p})} -
  \left(\frac{\Lambda_0}{\Lambda}\right)^\eta \frac{1 - K
    \left(\bar{p} \frac{\Lambda}{\Lambda_0}\right)}{K \left(\bar{p}
      \frac{\Lambda}{\Lambda_0}\right)} \rb\,.\label{twopoint}
\end{align}
The connected parts of the higher point functions are simply related by
the same change of normalization as
\begin{equation}
\vev{\bar{\phi} (\bar{p}_1) \cdots \bar{\phi}
  (\bar{p}_n)}^{\mathrm{conn}}_{\bar{S}_\tau} =
\left( \left(\frac{\Lambda_0}{\Lambda}\right)^\eta
  \Lambda^{D+2}\right)^{\frac{n}{2}} \prod_{i=1}^n \frac{K
  (\bar{p}_i)}{K\left( \bar{p}_i \frac{\Lambda}{\Lambda_0}\right)}
\cdot
\vev{\phi (\bar{p}_1 \Lambda) \cdots \phi (\bar{p}_n \Lambda)
}_{\Sb}^{\mathrm{conn}}\,,\label{higherpoint}
\end{equation}
where $n \ne 2$.  (These results are well known in the ERG literature.
See, for example, the review article~\cite{Igarashi:2009tj}.)

To understand the above results better, let us introduce a
$\Lambda$-dependent field by
\begin{equation}
\varphi (t, p) \equiv \frac{K (p/\Lambda)}{K(p/\Lambda_0)}\, \phi
(p)\,, 
\end{equation}
where the diffusion time $t$ is given by
\begin{equation}
t \equiv \frac{1}{\Lambda^2} - \frac{1}{\Lambda_0^2} =
\frac{e^{2 \tau}}{\mu^2} - \frac{1}{\Lambda_0^2}\,.\label{diffusion-t}
\end{equation}
Using the explicit form (\ref{gaussian}) of the cutoff function, we
obtain
\begin{equation}
\varphi (t, p) = e^{- t p^2} \phi (p)\,.\label{varphi-p}
\end{equation}
In coordinate space, this gives
\begin{equation}
\varphi (t, x) \equiv \int_p e^{i p x} \varphi (t,p)\,,\label{varphi-x}
\end{equation}
which satisfies the diffusion equation
\begin{equation}
\left( \partial_t - \partial^2 \right) \varphi (t,x) = 0\,,
\end{equation}
and the initial condition
\begin{equation}
\varphi (0, x) = \phi (x) \equiv \int_p e^{i p x} \phi (p)\,.
\end{equation}

Integrating (\ref{twopoint}, \ref{higherpoint}) over the momenta, and
using the notation $\varphi$, we obtain
\begin{align}
  \vev{\bar{\phi}(x)^2}_{\bar{S}_\tau} &=
  \left(\frac{\Lambda_0}{\Lambda}\right)^\eta \Lambda^{2-D}
  \vev{\varphi (t,x)^2}_{\Sb} + \int_{\bar{p}}
  \frac{K(\bar{p})\left(1 - K(\bar{p})\right)}{\bar{p}^2}
  \,,\label{twopoint-x}\\ 
  \vev{\bar{\phi}(x)^n}^{\mathrm{conn}}_{\bar{S}_\tau} &=
  \left(\left(\frac{\Lambda_0}{\Lambda}\right)^\eta \Lambda^{2-D}
  \right)^{\frac{n}{2}} \vev{\varphi (t,
    x)^n}_{\Sb}\,,\label{higherpoint-x}
\end{align}
where, assuming $\Lambda \ll \Lambda_0$ and $\eta < 2$, we have set
$K(\bar{p} \Lambda/\Lambda_0) = 1$ in evaluating the constant shift in
the two-point function.  The above gives an explicit relation between
the expectation values of the diffused fields with the bare action and
those of the elementary fields with the Wilson action.

Using (\ref{twopoint-x}) and (\ref{higherpoint-x}), we obtain, for
example,
\begin{equation}
  \frac{\vev{\varphi (t)^4}_{\Sb}^{\mathrm{conn}}}{\vev{\varphi (t)^2}_{\Sb}^2}
  = \frac{\vev{\varphi (t)^4}_{\Sb}}{\vev{\varphi (t)^2}_{\Sb}^2} - 3
  = \frac{\vev{\bar{\phi}^4}^{\mathrm{conn}}_{\bar{S}_\tau}} {\left(
      \vev{\bar{\phi}^2}_{\bar{S}_\tau} - \int_{\bar{p}} \frac{K(\bar{p})\left(1 -
          K(\bar{p})\right)}{\bar{p}^2}\right)^2}\,.
\end{equation}
Similarly, we obtain
\begin{equation}
\frac{\frac{1}{\Lambda^2} \vev{ \partial_\mu \varphi
    (t,x) \partial_\mu \varphi (t,x)}_{\Sb}}{\vev{\varphi
    (t)^2}_{\Sb}} = \frac{\vev{\partial_\mu \bar{\phi}
    (x) \partial_\mu \bar{\phi} (x)}_{\bar{S}_\tau} - \int_{\bar{p}}
  K(\bar{p}) \left(1 - K(\bar{p})\right)}{
  \vev{\bar{\phi}^2}_{\bar{S}_\tau} - \int_{\bar{p}}
  \frac{K(\bar{p})\left(1 - K(\bar{p})\right)}{\bar{p}^2} }\,.
\end{equation}
Suppose the bare theory $\Sb$ is critical so that $\bar{S}_\tau$
approaches a fixed point as $\tau \to \infty$ (hence $t \to \infty$):
\begin{equation}
\lim_{\tau \to \infty} \bar{S}_\tau = \bar{S}^*\,.
\end{equation}
We then obtain
\begin{align}
  \frac{\vev{\varphi (t)^4}_{\Sb}}{\vev{\varphi (t)^2}_{\Sb}^2} - 3
  &\overset{t \to \infty}{\longrightarrow}
  \frac{\vev{\bar{\phi}^4}^{\mathrm{conn}}_{\bar{S}^*}} {\left(
      \vev{\bar{\phi}^2}_{\bar{S}^*} - \int_{\bar{p}}
      \frac{K(\bar{p})\left(1 -
          K(\bar{p})\right)}{\bar{p}^2}\right)^2}\,,\\
  \frac{\frac{1}{\Lambda^2} \vev{ \partial_\mu \varphi
      (t,x) \partial_\mu \varphi (t,x)}_{\Sb}}{\vev{\varphi
      (t)^2}_{\Sb}} &\overset{t \to \infty}{\longrightarrow}
  \frac{\vev{\partial_\mu \bar{\phi} (x) \partial_\mu \bar{\phi}
      (x)}_{\bar{S}^*} - \int_{\bar{p}} K(\bar{p}) \left(1 -
      K(\bar{p})\right)}{ \vev{\bar{\phi}^2}_{\bar{S}^*} -
    \int_{\bar{p}} \frac{K(\bar{p})\left(1 -
        K(\bar{p})\right)}{\bar{p}^2} }\,.
\end{align}
The large $t$ behavior of the left-hand sides have been calculated
explicitly in the large $N$ limit of the $O(N)$ linear sigma model in
$D=3$~\cite{Aoki:2016ohw,Makino:2018rys}.

\section{Gradient flow for a renormalizable theory}

We next consider a bare action $\Sb$ that corresponds to a
renormalizable theory.  To discuss renormalization non-perturbatively,
we need to construct a renormalized trajectory $\tilde{S}_\tau$ that
can be traced back to the fixed point $\bar{S}^*$ under the ERG flow:
\begin{equation}
\lim_{\tau \to - \infty} \tilde{S}_\tau = \bar{S}^*\,.
\end{equation}
Let us outline the construction of the renormalized trajectory,
following Sec. 12 of~\cite{Wilson:1973jj}.

Given a bare action $\Sb [\phi]$ with momentum cutoff
$\Lambda_0$, let
\begin{equation}
\bar{S}_{\mathrm{bare}}  [\bar{\phi}] \equiv \Sb [\phi]
\end{equation}
be the corresponding action for the dimensionless field
\begin{equation}
\bar{\phi} (\bar{p}) = \Lambda_0^{\frac{D+2}{2}} \phi (\bar{p} \Lambda_0)\,.
\end{equation}
We can take the dimensionless squared mass $\bar{m}^2$ as the free
parameter of the bare action
$\bar{S}_{\mathrm{bare}} (\bar{m}^2) [\bar{\phi}]$.  We assume that
the theory is critical at
\begin{equation}
\bar{m}^2 = \bar{m}_{cr}^2\,.
\end{equation}
This means that the solution $\bar{S}_\tau$ of the ERG differential equation
(\ref{ERGdiffeq-Sbar}) with the initial condition
\begin{equation}
\bar{S}_{\tau=0} = \bar{S}_{\mathrm{bare}} (\bar{m}_{cr}^2)
\end{equation}
satisfies
\begin{equation}
\lim_{\tau \to \infty} \bar{S}_\tau = \bar{S}^*\,.
\end{equation}

We assume that the fixed point $\bar{S}^*$ has only one relevant
direction with scaling dimension $y > 0$.  (Otherwise, we need to tune
more than $\bar{m}^2$.)  Let $\bar{S}_\tau (g, \Lambda_0/\mu)$ be the
solution of (\ref{ERGdiffeq-Sbar}) satisfying the initial condition
\begin{equation}
\bar{S}_{\tau = 0} (g, \Lambda_0/\mu) = \bar{S}_{\mathrm{bare}}
\left(\bar{m}^2 (g, \Lambda_0/\mu)\right) \,,
\end{equation}
where
\begin{equation}
  \bar{m}^2 (g, \Lambda_0/\mu) = \bar{m}_{cr}^2 + g
  \left(\frac{\mu}{\Lambda_0}\right)^y\,. 
\end{equation}
$\mu$ is an arbitrary reference momentum scale where the parameter
$g$ is defined.  Note that
$\bar{m}^2 (g, \Lambda_0/\mu)$ satisfies
\begin{equation}
\left( y g \frac{\partial}{\partial g} + \Lambda_0
  \frac{\partial}{\partial \Lambda_0} \right) \bar{m}^2 (g,
\Lambda_0/\mu) = 0\,.
\end{equation}
This implies
\begin{equation}
\left( y g \frac{\partial}{\partial g} + \Lambda_0
  \frac{\partial}{\partial \Lambda_0} \right) \bar{S}_{\tau = 0} (g,
\Lambda_0/\mu) = 0\,.\label{g-Lambdazero}
\end{equation}

We can then define a renormalized trajectory by the limit
\begin{equation}
\tilde{S} (g) \equiv \lim_{\Lambda_0 \to \infty}
\bar{S}_{\ln \frac{\Lambda_0}{\mu}} (g, \Lambda_0/\mu)\,.\label{def-Stilde}
\end{equation}
For the limit to exist, we must find
\begin{equation}
\Lambda_0 \frac{\partial}{\partial \Lambda_0} \bar{S}_{\ln
  \frac{\Lambda_0}{\mu}} (g, \Lambda_0/\mu) \overset{\Lambda_0 \to
  \infty}{\longrightarrow} 0\,.\label{condition}
\end{equation}
For an explanation that such a limit exists, we refer the reader to
standard references such as Sec. 12 of~\cite{Wilson:1973jj}.
Since (\ref{g-Lambdazero}) implies
\begin{equation}
\left( y g \frac{\partial}{\partial g} + \Lambda_0
  \frac{\partial}{\partial \Lambda_0} \right) \bar{S}_\tau (g,
\Lambda_0/\mu) = 0
\end{equation}
for any $\tau > 0$, we obtain, from (\ref{condition}), 
\begin{equation}
y g \frac{\partial}{\partial g} \tilde{S} (g)
= \lim_{\Lambda_0 \to \infty} \left( \Lambda_0
  \frac{\partial}{\partial \Lambda_0} \bar{S}_{\ln \frac{\Lambda_0}{\mu}}
  (g, \Lambda/\mu) \right)_{\Lambda = \Lambda_0}\,.
\end{equation}
Hence, from (\ref{ERGdiffeq-Sbar}), $\tilde{S} (g)$ satisfies the ERG
differential equation
\begin{align}
y g \frac{\partial}{\partial g} e^{\tilde{S} (g) [\bar{\phi}]}
&= \int_p \left[ \left( \frac{\Delta (p)}{K(p)} + \frac{D+2}{2} -
    \frac{\eta}{2} + p \cdot \partial_p \right) \bar{\phi} (p)
  \frac{\delta}{\delta 
    \bar{\phi} (p)} \right.\notag\\
&\quad \left. + \frac{\Delta (p) - \eta K(p)\left(1 -
  K(p)\right)}{p^2} \frac{1}{2} \frac{\delta^2}{\delta \bar{\phi} (p) \delta
  \bar{\phi} (-p)} \right]\, e^{\tilde{S} (g)
[\bar{\phi}]}\,,\label{ERGdiffeq-Stilde} 
\end{align}
where we have omitted the bar over the dimensionless momentum $p$ to
simplify the notation.
\begin{figure}[hbt]
\centering
\includegraphics{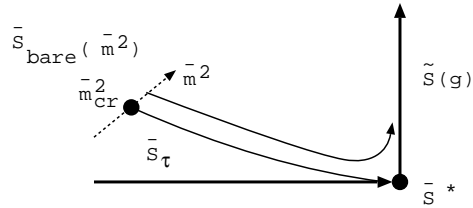}
\caption{ERG flows $\bar{S}_\tau$ and a renormalized trajectory $\tilde{S} (g)$}
\end{figure}

Now that we have constructed a renormalized trajectory $\tilde{S} (g)$,
let us rewrite (\ref{def-Stilde}) as the relation between bare and
renormalized Wilson actions of dimensionful fields.  We first define a
bare action with cutoff $\Lambda_0$ by
\begin{equation}
\Sb (g, \Lambda_0/\mu) [\phi] \equiv \bar{S}_{\mathrm{bare}} \left(
  \bar{m}^2 (g, \Lambda_0/\mu) \right) [\bar{\phi}]\,,
\end{equation}
where
\begin{equation}
\phi (p) \equiv \Lambda_0^{-\frac{D+2}{2}} \bar{\phi}
(p/\Lambda_0)\,.\label{phi-bare} 
\end{equation}
The squared mass of $\Sb (g,\Lambda_0/\mu)$ is given by
\begin{equation}
\Lambda_0^2\, \bar{m}^2 (g, \Lambda_0/\mu) = \Lambda_0^2\, \bar{m}_{cr}^2
+ g \Lambda_0^2 \left(\frac{\mu}{\Lambda_0}\right)^y\,.
\end{equation}
We then define a renormalized Wilson action with cutoff $\Lambda$ by
\begin{equation}
  \SL [\phi] \equiv \tilde{S} (g_\Lambda) [\bar{\phi}]\,,\label{SL-renormalized}
\end{equation}
where 
\begin{equation}
\phi (p) \equiv \Lambda^{-\frac{D+2}{2}} \bar{\phi}
(p/\Lambda)\,,\label{phi-renormalized} 
\end{equation}
and the dimensionless parameter $g_\Lambda$ is defined by 
\begin{equation}
g_\Lambda \equiv g \left(\frac{\mu}{\Lambda}\right)^y\,.\label{gLambda}
\end{equation}
Since 
\begin{equation}
\bar{m}^2 (g_\Lambda, \Lambda_0/\Lambda)  = \bar{m}^2 (g,
\Lambda_0/\mu)\,,\label{mbar-relation} 
\end{equation}
(\ref{def-Stilde}) gives
\begin{align}
  \tilde{S} (g_\Lambda) &= \lim_{\Lambda_0 \to \infty} \bar{S}_{\ln
    \frac{\Lambda_0}{\mu}} (g_\Lambda,
  \Lambda_0/\mu)\notag\\
  &= \lim_{\Lambda_0 \to \infty} \bar{S}_{\ln
    \frac{\Lambda_0}{\Lambda}}
  (g_\Lambda, \Lambda_0/\Lambda)\notag\\
  &= \lim_{\Lambda_0 \to \infty} \bar{S}_{\ln
    \frac{\Lambda_0}{\Lambda}} (g, \Lambda_0/\mu)\,.
\end{align}
Since $\SL$ is $\tilde{S} (g_\Lambda)$ for the dimensionful field
(\ref{phi-renormalized}), and $\Sb (g,\Lambda_0/\mu)$ is
$ \bar{S}_{\mathrm{bare}} \left(\bar{m}^2 (g,\Lambda_0/\mu)\right)=
\bar{S}_{\tau=0} (g,\Lambda_0/\mu)$
for (\ref{phi-bare}), we find that $\SL$ is obtained from the bare
action $\Sb (g, \Lambda_0/\mu)$ by solving the ERG differential
equation (\ref{ERGdiffeq}) from $\Lambda_0$ to $\Lambda$ (and taking
the limit $\Lambda_0 \to \infty$).  Hence, $\SL$ and
$\Sb (g,\Lambda_0/\mu)$ are related by (\ref{integral}) as
\begin{align}
  &e^{\SL [\phi]} = \int [d\phi'] \exp \left[ \Sb (g, \Lambda_0/\mu)
    [\phi'] - \frac{1}{2} \int_p
    \frac{p^2}{\left(\frac{\Lambda_0}{\mu}\right)^\eta
      \frac{1-K(p/\Lambda)}{K(p/\Lambda)} -
      \left(\frac{\Lambda_0}{\mu}\right)^\eta
      \frac{1-K(p/\Lambda_0)}{K(p/\Lambda_0}}
  \right.\notag\\
  &\left. \quad\times \left(
      \left(\frac{\Lambda_0}{\mu}\right)^{\frac{\eta}{2}} \frac{\phi'
        (p)}{K(p/\Lambda_0)} -
      \left(\frac{\Lambda}{\mu}\right)^{\frac{\eta}{2}} \frac{\phi
        (p)}{K(p/\Lambda)} \right) \left(
      \left(\frac{\Lambda_0}{\mu}\right)^{\frac{\eta}{2}} \frac{\phi'
        (-p)}{K(p/\Lambda_0)} -
      \left(\frac{\Lambda}{\mu}\right)^{\frac{\eta}{2}} \frac{\phi
        (-p)}{K(p/\Lambda)} \right) \right]\,.
\end{align}
This implies that the correlation functions are related by
\begin{align}
  \vev{\phi (p) \phi (q)}_{\SL} &=
  \left(\frac{K(p/\Lambda)}{K(p/\Lambda_0)}\right)^2
  \left(\frac{\Lambda_0}{\Lambda}\right)^\eta \vev{\phi (p) \phi
    (q)}_{\Sb
    (g,\Lambda_0/\mu)}\notag\\
  &\, + \delta (p+q) \frac{K(p/\Lambda)^2}{p^2} \lb \frac{1 -
    K(p/\Lambda)}{K(p/\Lambda)} -
  \left(\frac{\Lambda_0}{\Lambda}\right)^\eta \frac{1 -
    K(p/\Lambda_0)}{K(p/\Lambda_0)}\rb \notag\\
  &\overset{\Lambda_0 \to \infty}{\longrightarrow}
  \left(\frac{\Lambda_0}{\Lambda}\right)^\eta \vev{\varphi (t, p)
    \varphi
    (t, q)}_{\Sb (g,\Lambda_0/\mu)} \notag\\
  &\qquad\qquad + \delta (p+q) \frac{K(p/\Lambda) \left(1 -
      K(p/\Lambda)\right)}{p^2}\,,\label{twopoint-SL-Sbare}
\end{align}
and
\begin{align}
\vev{\phi (p_1) \cdots \phi (p_n)}^{\mathrm{conn}}_{\SL} 
&= \left(\frac{\Lambda_0}{\Lambda}\right)^{\frac{n}{2} \eta} 
\prod_{i=1}^n \frac{K(p_i/\Lambda)}{K(p_i/\Lambda_0)} \cdot
\vev{\phi (p_1) \cdots \phi (p_n)}^{\mathrm{conn}}_{\Sb (g,
  \Lambda_0/\mu)}\notag\\
&\overset{\Lambda_0 \to \infty}{\longrightarrow}
  \left(\frac{\Lambda_0}{\Lambda}\right)^{\frac{n}{2} \eta} 
 \vev{\varphi (t, p_1) \cdots \varphi (t, p_n)}^{\mathrm{conn}}_{\Sb (g,
  \Lambda_0/\mu)}\,,\label{higherpoint-SL-Sbare}
\end{align}
where the diffused field $\varphi (t, p)$ is given by
(\ref{diffusion-t}) and (\ref{varphi-p}).  Integrating over the
momenta and taking $\Lambda_0 \to \infty$, we obtain the expectation
values of local products:
\begin{align}
  \vev{\phi (x)^2}_{\SL} &= \lim_{\Lambda_0 \to \infty}
  \left(\frac{\Lambda_0}{\Lambda}\right)^\eta \vev{\varphi
    (t,x)^2}_{\Sb (g,\Lambda_0/\mu)} + \Lambda^{D-2} \int_p \frac{K(p)
    \left(1 -
      K(p)\right)}{p^2}\,,\label{twopoint-x-Lambda}\\
  \vev{\phi (x)^n}^{\mathrm{conn}}_{\SL} &= \lim_{\Lambda_0 \to
    \infty} \left(\frac{\Lambda_0}{\Lambda}\right)^{n \frac{\eta}{2}}
  \vev{\varphi (t,x)^n}^{\mathrm{conn}}_{\Sb
    (g,\Lambda_0/\mu)}\,.\label{higherpoint-x-Lambda}
\end{align}
Note that the diffused field only needs the standard wave function
renormalization in the continuum limit $\Lambda_0 \to \infty$.  Local
products of $\phi (x)$ have no short distance singularities thanks to
the momentum cutoff $\Lambda$.  (\ref{twopoint-x-Lambda},
\ref{higherpoint-x-Lambda}) give the concrete correspondence between
the gradient flow ($t$) and RG flow ($\Lambda$) for the renormalized
theory.

Before closing this section, we would like to relate the correlation
functions in the continuum limit to those obtained by the Wilson
action $\SL$.  From (\ref{twopoint-SL-Sbare}) and
(\ref{higherpoint-SL-Sbare}) we obtain
\begin{align}
\lim_{\Lambda_0 \to \infty} \left(\frac{\Lambda_0}{\mu}\right)^\eta
\vev{\phi (p) \phi (q)}_{\Sb (g, \Lambda_0/\mu)}
&= \left(\frac{\Lambda}{\mu}\right)^\eta \vvev{\phi (p) \phi
  (q)}_{\Lambda} \,,\label{twopoint-limit}\\
\lim_{\Lambda_0 \to \infty}
\left(\frac{\Lambda_0}{\mu}\right)^{\frac{\eta}{2} n} \vev{\phi (p_1)
  \cdots \phi (p_n)}^{\mathrm{conn}}_{\Sb (g, \Lambda_0/\mu)}
&= \left(\frac{\Lambda}{\mu}\right)^{\frac{\eta}{2} n} \vvev{\phi
  (p_1) \cdots \phi (p_n)}_{\Lambda}^{\mathrm{conn}} \,,\label{higherpoint-limit}
\end{align}
where we define
\begin{align}
\vvev{\phi (p) \phi (q)}_{\Lambda} &\equiv 
\frac{1}{K(p/\Lambda)^2} \left( \vev{\phi (p)\phi (q)}_{\SL} -
  \frac{K(p/\Lambda) \left(1 - K(p/\Lambda)\right)}{p^2} \delta (p+q)
\right)\,,\label{twopoint-mod-Lambda}\\
\vvev{\phi (p_1) \cdots \phi (p_n)}_{\Lambda}^{\mathrm{conn}}
&\equiv  \prod_{i=1}^n
\frac{1}{K(p_i/\Lambda)} \cdot \vev{\phi (p_1) \cdots \phi
  (p_n)}^{\mathrm{conn}}_{\SL}\,.\label{higherpoint-mod-Lambda}
\end{align}
The field of the Wilson action corresponds to a diffused field of the
continuum limit, and we use the factor $1/K(p/\Lambda)$ for each
$\phi (p)$ to reverse diffusion.  Thus, using a Wilson action with a
finite cutoff $\Lambda$, we manage to construct the correlation
functions in the continuum limit, valid for any momenta.

The correlation functions with double brackets are the continuum limit
defined at renormalization scale $\Lambda$.
They satisfy the RG equation with anomalous dimension $\frac{\eta}{2}$:
\begin{align}
\vvev{\phi (p) \phi (q)}_{\Lambda'} &=
\left(\frac{\Lambda}{\Lambda'}\right)^\eta \vvev{\phi (p) \phi
  (q)}_{\Lambda}\,,\\
\vvev{\phi (p_1) \cdots \phi (p_n)}_{\Lambda'}^{\mathrm{conn}} &=
\left(\frac{\Lambda}{\Lambda'}\right)^{\frac{\eta}{2} n} \vvev{\phi
  (p_1) \cdots \phi (p_n)}_{\Lambda}\,.
\end{align}
This explains the powers of $\Lambda/\mu$, necessary to make the
right-hand sides of (\ref{twopoint-limit}) and
(\ref{higherpoint-limit}) independent of $\Lambda$.

\section{The small time expansions}

In the previous section we obtained the relation
(\ref{twopoint-x-Lambda}) \& (\ref{higherpoint-x-Lambda}) between the
expectation value of $\varphi (t,x)^n$ in the continuum limit and that
of $\phi (x)^n$ with the Wilson action $\SL$.  We now wish to
understand the behavior of the latter as $\Lambda \to \infty$,
or equivalently $t \to 0$.  In particular we wish to derive small
$t$ expansions analogous to those obtained for QCD in~\cite{Luscher:2011bx}.

By construction (\ref{SL-renormalized}, \ref{phi-renormalized}), we
obtain
\begin{align}
  \vev{\phi (p) \phi (q)}_{\SL} &= \Lambda^{- (D+2)} \vev{\bar{\phi}
    (p/\Lambda) \bar{\phi} (q/\Lambda)}_{\tilde{S}
    (g_\Lambda)}\,,\\
  \vev{\phi (p_1) \cdots \phi (p_n)}^{\mathrm{conn}}_{\SL} &=
  \Lambda^{- (D+2) \frac{n}{2}} \vev{\bar{\phi} (p_1/\Lambda) \cdots
    \bar{\phi} (p_n/\Lambda)}^{\mathrm{conn}}_{\tilde{S}
    (g_\Lambda)}\,.
\end{align}
Hence, integrating over the momenta, we obtain
\begin{align}
\vev{\phi^2}_{\SL} &= \Lambda^{D-2} \vev{\bar{\phi}^2}_{\tilde{S}
                     (g_\Lambda)}\,,\label{twopoint-SL-Stilde}\\
\vev{\phi^n}^{\mathrm{conn}}_{\SL} &= \Lambda^{(D-2)\frac{n}{2}}
                     \vev{\bar{\phi}^n}^{\mathrm{conn}}_{\tilde{S}
                     (g_\Lambda)}\,.\label{higherpoint-SL-Stilde}
\end{align}
Let us introduce the dimensionless analogs of
(\ref{twopoint-mod-Lambda}) and (\ref{higherpoint-mod-Lambda}) by
\begin{align}
\vvev{\bar{\phi} (p) \bar{\phi} (q)}_{g_\Lambda} 
&\equiv \frac{1}{K(p)^2} \left( \vev{\bar{\phi} (p) \bar{\phi}
  (q)}_{\tilde{S} (g_\Lambda)} - \frac{K(p)\left(1 - K(p)\right)}{p^2}
  \delta (p+q) \right)\,,\\
\vvev{\bar{\phi} (p_1) \cdots \bar{\phi} (p_n)}^{\mathrm{conn}}_{g_\Lambda} 
&\equiv \prod_{i=1}^n \frac{1}{K(p_i)}\cdot \vev{\bar{\phi} (p_1)
  \cdots \bar{\phi} (p_n)}^{\mathrm{conn}}_{\tilde{S} (g_\Lambda)}\,.
\end{align}
These satisfy the scaling laws:
\begin{align} 
\vvev{\bar{\phi} (p e^\tau) \bar{\phi} (q e^\tau)}_{g_\Lambda e^{y
    \tau}} &= e^{- (D+2-\eta) \tau} \vvev{\bar{\phi} (p) \bar{\phi}
  (q)}_{g_\Lambda}\,,\\
\vvev{\bar{\phi} (p_1 e^\tau) \cdots \bar{\phi} (p_n
  e^\tau)}_{g_\Lambda e^{y \tau}}^{\mathrm{conn}} &= e^{- \frac{n}{2}
  (D+2-\eta) \tau} \vvev{\bar{\phi} (p_1) \cdots \bar{\phi}
  (p_n)}_{g_\Lambda}^{\mathrm{conn}}\,.
\end{align} 
Correspondingly, the correlation functions in coordinate
space, defined by
\begin{align} \vvev{\bar{\phi} (x) \bar{\phi} (0)}_{g_\Lambda} &\equiv
\int_p e^{i p x} \int_q \vvev{\bar{\phi} (p) \bar{\phi}
(q)}_{g_\Lambda}\,,\\ \vvev{\bar{\phi} (x_1) \cdots \bar{\phi}
(x_{n-1}) \bar{\phi} (0)}_{g_\Lambda}^{\mathrm{conn}} &\equiv
\prod_{i=1}^{n-1} \int_{p_i} e^{i p_i x_i} \int_{p_n} \vvev{\bar{\phi} (p_1)
\cdots \bar{\phi} (p_n)}_{g_\Lambda}\,,
\end{align} 
satisfy the scaling laws:
\begin{align} 
\vvev{\bar{\phi} (x e^{-\tau}) \bar{\phi} (0)}_{g_\Lambda e^{y \tau}}
&= e^{(D-2+\eta) \tau} \vvev{\bar{\phi} (x) \bar{\phi}
  (0)}_{g_\Lambda}\,,\\
\vvev{\bar{\phi} (x_1 e^{-\tau}) \cdots \bar{\phi} (x_n
  e^{-\tau})}_{g_\Lambda e^{y \tau}} &= e^{\frac{n}{2} (D-2+\eta)
  \tau} \vvev{\bar{\phi} (x_1) \cdots \bar{\phi}
  (x_n)}_{g_\Lambda}^{\mathrm{conn}}\,.
\end{align}

Thus, we obtain
\begin{align} \vev{\bar{\phi}^2}_{\tilde{S} (g_\Lambda)} &= \int_{p,
q}\vev{\bar{\phi} (p) \bar{\phi} (q)}_{\tilde{S} (g_\Lambda)}\notag\\
&= \int_{p,q} K(p)^2 \vvev{\bar{\phi} (p) \bar{\phi} (q)}_{g_\Lambda}
+ \int_p \frac{K(p)\left(1 - K(p)\right)}{p^2}\notag\\ &= \int d^D x
\, \mathcal{K}_2 (x) \vvev{\bar{\phi} (x) \bar{\phi} (0)}_{g_\Lambda}
+ \int_p \frac{K(p)\left(1 - K(p)\right)}{p^2}\,,
\end{align}
and
\begin{align}
\vev{\bar{\phi}^n}^{\mathrm{conn}}_{\tilde{S} (g_\Lambda)} &=
\int_{p_1, \cdots, p_n} \vev{\bar{\phi} (p_1) \cdots \bar{\phi}
(p_n)}^{\mathrm{conn}}_{\tilde{S} (g_\Lambda)}\notag\\ &=
\int_{p_1,\cdots,p_n} \prod_{i=1}^n K(p_i)\, \vvev{\bar{\phi} (p_1)
\cdots \bar{\phi} (p_n)}_{g_\Lambda}\notag\\ &= \int d^D x_1 \cdots
d^D x_{n-1}\, \mathcal{K}_n (x_1,\cdots,x_{n-1})\, \vvev{\bar{\phi}
(x_1) \cdots \bar{\phi} (x_{n-1}) \bar{\phi} (0)}_{g_\Lambda}\,,
\end{align} 
where we have defined
\begin{align} 
  \mathcal{K}_2 (x) &\equiv \int_p e^{i p x} K(p)^2 = \frac{1}{\left(2
      \sqrt{2 \pi}\right)^D} \exp \left( - \frac{1}{8} x^2 \right)\,,\\
  \mathcal{K}_n (x_1, \cdots, x_{n-1}) &\equiv \int_{p_1, \cdots,
    p_{n-1}} e^{i \sum_{i=1}^{n-1} p_i x_i} \prod_{i=1}^{n-1} K(p_i) \cdot K\left( - p_1
    - \cdots - p_{n-1}\right)\\
&= \frac{1}{\left(2^{n-1} \pi^{\frac{n-1}{2}} \,\sqrt{n}\right)^D} \exp
\left[ - \frac{1}{4} \sum_{i=1}^{n-1} x_i^2 + \frac{1}{4 n} (x_1 +
  \cdots +x_{n-1})^2 \right]\,.\notag
\end{align}
The functions $\mathcal{K}$'s are gaussian with a range of order $1$
in coordinate space.

Since the mass scale of $\tilde{S} (g_\Lambda)$ is of order
$g_\Lambda^{\frac{1}{y}}$, the distance of order $1$ is very short
compared with the inverse mass $g_\Lambda^{- \frac{1}{y}}$ as long as
$g_\Lambda \ll 1$.  Because $g_\Lambda$ is given by (\ref{gLambda}),
we obtain $g_\Lambda \ll 1$ if we take
\begin{equation}
\Lambda \gg \mu \, g^{\frac{1}{y}}\quad\textrm{or equivalently}\quad t =
\frac{1}{\Lambda^2} \ll \frac{g^{-\frac{2}{y}}}{\mu^2}\,.
\end{equation}
Hence, for such large $\Lambda$ and the
coordinates of order $1$, we can use the short distance expansions:
\begin{align}
  \vvev{\bar{\phi} (x) \bar{\phi} (0)}_{g_\Lambda} 
  &= \sum_i C_{2, i} (g_\Lambda; x) \vev{\Op_i (0)}_{\tilde{S}(g_\Lambda)} \,,\label{ope-two}\\
\vvev{\bar{\phi} (x_1) \cdots \bar{\phi} (x_{n-1}) \bar{\phi}
  (0)}^{\mathrm{conn}}_{g_\Lambda} 
&= \sum_i C_{n, i} (g_\Lambda; x_1,\cdots, x_{n-1}) \vev{\Op_i
  (0)}_{\tilde{S}(g_\Lambda)} \,, \label{ope-higher}
\end{align}
where $\Op_i$ is a local composite operator of scale dimension $D -
y_i$ whose expectation values are given by
\begin{equation}
\vev{\Op_i}_{\tilde{S} (g_\Lambda)} = g_\Lambda^{\frac{D-y_i}{y}}\,.
\end{equation}
The coefficient functions satisfy the RG equations:
\begin{align}
\left( y g_\Lambda \partial_{g_\Lambda} - x \cdot \partial_x -
  (D-2+\eta - (D-y_i))\right) C_{2,i} (g_\Lambda; x) &=0\,,\\
\left(  y g_\Lambda \partial_{g_\Lambda} - \sum_{i=1}^{n-1} x_i
  \cdot \partial_{x_i} -
  \left( \frac{n}{2}(D-2+\eta) - (D-y_i)\right)\right) C_{n,i}
(g_\Lambda; x_1,\cdots,x_{n-1}) &=0\,.
\end{align}
For $x$'s of order $1$, the coefficient functions
$C_{n, i} (g_\Lambda; x_1, \cdots, x_{n-1})$ can be expanded in powers
of $g_\Lambda \ll 1$.  Hence,
\begin{align}
\mathcal{C}_{2,i} (g_\Lambda) &\equiv \int d^D x\, \mathcal{K}_2 (x)\,
C_{2,i} (g_\Lambda; x)\,,\\
\mathcal{C}_{n,i} (g_\Lambda) &\equiv \int d^D x_1 \cdots d^D x_{n-1}\,
\mathcal{K}_n (x_1,\cdots,x_{n-1})  \, C_{n,i} (g_\Lambda; x_1, \cdots,
  x_{n-1})
\end{align}
can be expanded in powers of $g_\Lambda$. 

We thus obtain the large $\Lambda$ expansions as
\begin{align}
\vev{\bar{\phi}^2}_{\tilde{S} (g_\Lambda)} 
&= \sum_i \vev{\Op_i}_{\tilde{S} (g_\Lambda)} \, \mathcal{C}_{2,i} (g_\Lambda)\,,\\
\vev{\bar{\phi}^n}^{\mathrm{conn}}_{\tilde{S} (g_\Lambda)} 
&= \sum_i   \vev{\Op_i}_{\tilde{S} (g_\Lambda)} \,\mathcal{C}_{n,i} (g_\Lambda)\,.
\end{align}
Using (\ref{twopoint-x-Lambda}, \ref{higherpoint-x-Lambda}) and
(\ref{twopoint-SL-Stilde}, \ref{higherpoint-SL-Stilde}), we can
rewrite the above for the continuum limit:
\begin{align}
\lim_{\Lambda_0 \to \infty} \left(\frac{\Lambda_0}{\mu}\right)^\eta
\vev{\varphi (t)^2}_{\Sb (g, \Lambda_0/\mu)}
&= \Lambda^{D-2} \left(\frac{\Lambda}{\mu}\right)^\eta \sum_i
\vev{\Op_i}_{\tilde{S} (g_\Lambda)} \,\mathcal{C}_{2,i} (g_\Lambda)\,,\\
\lim_{\Lambda_0 \to \infty} \left(\frac{\Lambda_0}{\mu}\right)^{n
  \frac{\eta}{2}} \vev{\varphi (t)^n}^{\mathrm{conn}}_{\Sb (g,
  \Lambda_0/\mu)} 
&= \Lambda^{(D-2)\frac{n}{2}} \left(\frac{\Lambda}{\mu}\right)^{n
  \frac{\eta}{2}} \sum_i \vev{\Op_i}_{\tilde{S} (g_\Lambda)}
\,\mathcal{C}_{n,i} (g_\Lambda)\,.
\end{align}
This is the analog of the small $t$ expansions obtained for QCD
in~\cite{Luscher:2011bx}.  Here, we have derived them by relating them to
the standard short distance expansions (\ref{ope-two}, \ref{ope-higher}).

\section{Conclusions}

In this paper we have considered the gradient flow of a real scalar
field obeying the simple diffusion equation without potential terms.
We have then shown that the correlation functions of diffused fields
match with those of elementary fields of a Wilson action that has a
finite momentum cutoff.  We have only discussed formalism, and we plan
to provide concrete examples of the correspondence in a future
publication.

Obviously we have scratched only the tip of an iceberg.  In theories
such as gauge theories and non-linear sigma models, the fields are
continuous but live naturally in a compact space, and the diffusion
equations that respect the geometry of the compact space should be and
have been introduced~\cite{Luscher:2010iy,Makino:2014sta}. Both gauge
theories and non-linear sigma models can be formulated in ERG, but the
realization of symmetry is not manifest (see~\cite{Igarashi:2009tj}
for example).  The exact manner of the correspondence between the gradient
flow and RG flow is not be obvious, but we would be surprised if there
were not any.

\section*{Acknowledgment}

The work of H.~Suzuki is supported in part by JSPS Grant-in-Aid for
Scientific Research Grant Number JP16H03982.

\end{document}